# A high luminosity ERL on ring e⁻e⁺ collider for a super charm factory

E. Recepoglu[1], S. Sultansoy[2,3]

[1]*Sarayköy Nuclear Research and Training Center (SNRTC), 06983, Kazan, Ankara, Turkey*

[2]*TOBB University of Economics and Technology, Physics Department, 06560, Sogutozu, Ankara, Turkey*

[3]*Institute of Physics, Academy of Sciences, H.Cavid Avenue 33, Baku, Azerbaijan.*

A high luminosity energy recovery linac on ring type electron-positron collider operating as super charm factory is proposed. It is shown that the luminosity $L = 10^{35} \text{cm}^{-2}\text{s}^{-1}$ and more can be achieved for center of mass energy $\sqrt{s} = 3.77$ GeV. The physics goals of this machine in investigation for charmed particles properties are briefly discussed.

**I. INTRODUCTION**

An idea of colliding of the electron beam from linac with a beam stored in a ring [1] is widely discussed during the last decades with two purposes:

- to achive the TeV energy scale in lepton-hadron and photon-hadron collisions (see review articles [2] and references therein),
- to construct high luminosity particle factories, namely , B-factory [3], φ-factory [4,5] c−τ-factory [6] etc.

Today, linac-ring type B-factory has lost its attractiveness with KEK-B [7] and PEP-B [8] colliders under operation and, especially, Super B proposals [9]. In addition, Super-B factories will copiously produce $\tau$ leptons (the moderate decreasing of the $\tau$ pair production cross section at $\sqrt{s} \approx 10$ GeV is compensated by high luminosity). As a result only a charm factory option of linac-ring type factories still preserves its actuality. Therefore, we inclined towards charm factory option. In order to search for charm mixing and CP violation by exploiting quantum coherence and to search for rare decays by using a background-free environment, unique opportunities is offered by $\Psi(3S)$. Therefore, the center of mass energy is fixed by the mass of $\Psi(3770)$ resonance. Existing CLEO-c [10] works with L = $10^{32}$ cm$^{-2}$s$^{-1}$. The BEPC charm factory [11] has design luminosity of $10^{33}$ cm$^{-2}$s$^{-1}$. Therefore, charm factory with L > $10^{34}$ cm$^{-2}$s$^{-1}$ will contribute charm physics greatly. It was shown in [12] that linac-ring option gives opportunity to achieve $L = 10^{34} \text{cm}^{-2}\text{s}^{-1}$. The main restriction on luminosity coming from linac beam power can be relaxed by using of energy recovery linac (ERL). In principle, ERL technology will give opportunity to construct super-charm factory with L well exceeding $10^{35}$ cm$^{-2}$s$^{-1}$. Linac-ring type charm factory is one of the four main parts of the TAC (Turkic Accelerator Complex) Project [13], which is developed since 1997 with the support of Turkish State Planning Organization and planned to be realized before 2020. Recently, a ring-ring tau-charm factory based on the crab waist collision with luminosity of $10^{35} \text{cm}^{-2}\text{s}^{-1}$ has been proposed at Novosibirsk Budker Institute of Nuclear Physics [14] and high intensity linear e⁻e⁺ collider for a tau-charm factory with same luminosity is discussed in [15]. Therefore, all three possible type colliders (ring-ring, linac-ring and linac-linac) are considered as super charm factory candidates.

Concerning the charm physics search program [16]:

- Even with $L = 10^{34} \text{cm}^{-2}\text{s}^{-1}$ there are a number of processes which will be better studied at dedicated charm factory than at the super-B factory with $L = 10^{36} \text{cm}^{-2}\text{s}^{-1}$
- With $L = 10^{35} \text{cm}^{-2}\text{s}^{-1}$ dedicated charm factory cover almost all topics which can be investigated at super-B
- With $L = 10^{36}$ cm$^{-2}$s$^{-1}$ super-charm factory will give opportunity to touch charm physics well further than super-B.

In this paper, a high luminosity ERL-ring electron-positron collider serving as super charm factory is studied. In section II and III we present general discussion of beam dynamics aspects of ERL-ring type colliders and proposed parameters of super charm factory, respectively. Short comments on the physics search potential of the proposed machine are given in section IV. In the final section, we give some concluding remarks.

## II. GENERAL CONSIDERATIONS

The center of mass energy is given by

$$\sqrt{s} = 2\sqrt{E_{e^+} E_{e^-}} \cos\theta \tag{1}$$

where $E_{e^-}$ is the energy of electrons accelerated in the ERL, $E_{e^+}$ is the energy of positrons stored in the ring and $\theta$ is the crossing angle. For charm factories, it is important to have $\Delta(\sqrt{s}) < \Gamma_{\Psi(3S)}$ in order to use the advantage of resonant production of $\Psi(3S)$ mesons: $m_{\Psi(3S)} = 3772.92 \pm 0.35$ MeV with $\Gamma_{\Psi(3S)} = 27.3 \pm 1.0$ MeV [17].

The luminosity of $e^-e^+$ collider keeping in mind crab crossing is given by

$$L = \frac{N^+ N^-}{4\pi\sigma_y \sqrt{(\sigma_z \tan\theta/2)^2 + \sigma_x^2}} f_c \tag{2}$$

where $\sigma_{x,y} = \sqrt{\beta_{x,y}\varepsilon_{x,y}}$ and $f_c$ is the collision frequency, $N^+$ and $N^-$ are the number of particles in the positron and electron bunches, respectively; $\sigma_x$ is the beam size in the horizontal, $\sigma_y$ is the beam size in the vertical and $\sigma_z$ is the beam size in the longitudinal direction; $\varepsilon$ is the beam emittance, $\beta$ is the beta function at the collision point in each plane and $\theta$ is the crossing angle between the beam lines at the interaction point (IP).

Another restriction comes from tune shift of positron beam due to interaction with electron bunches. The horizontal $\xi_x$ and the vertical $\xi_y$ tune shifts are given as following [18]:

$$\xi_x = \frac{r_e N^-}{2\pi\gamma^+} \frac{\beta_x}{\sigma_x^2\left[(1+\varphi^2)+\frac{\sigma_y}{\sigma_x}\sqrt{1+\varphi^2}\right]} \qquad (3)$$

$$\xi_y = \frac{r_e N^-}{2\pi\gamma^+} \frac{\beta_y}{\sigma_y(\sigma_x\sqrt{1+\varphi^2}+\sigma_y)} \qquad (4)$$

The Piwinski angle $\varphi$ is defined as:

$$\varphi = \frac{\sigma_z}{\sigma_x}\tan\frac{\theta}{2} \approx \frac{\sigma_z}{\sigma_x}\frac{\theta}{2} \qquad (5)$$

where $\theta$ is crossing angle, $\sigma_x$ is the horizontal rms bunch size, $\sigma_z$ is the rms bunch length, $N^-$ is the number of electrons per ERL bunch and $\gamma^+$ the Lorentz factor for the positrons in the ring.

The loss of particles because of scattering at the interaction point at a rate proportional to the machine luminosity has a crucial contribution to beam life time. The loss of particles due to QED process $e^+e^- \rightarrow e^+e^-\gamma$ (radiative Bhabha) and $e^+e^- \rightarrow e^+e^-$ (elastic Bhabha) that scatter beam particles outside the ring acceptance should be taken into account. The rate of loss for ring which depends on luminosity L and on cross section $\sigma_i = \sigma_{rad} + \sigma_{elastic}$ is given by

$$\frac{dN_i}{dt} = -\sigma_i L \qquad (6)$$

In our case $\sigma_{el} \ll \sigma_{rad}$ and therefore can be neglected. The radiative Bhabha process cross section of the loss of particles from beam is given with a good approximation as [19]:

$$\sigma_{rad} \approx \frac{16\alpha r_e^2}{3}\left[-\left(\ln\left(\frac{\Delta E}{E}\right)_{accept} + \frac{5}{8}\right)\left(\ln(4\gamma_{e^+}\gamma_{e^-}) - \frac{1}{2}\right) + \frac{1}{2}\ln^2\left(\frac{\Delta E}{E}\right)_{accept} - \frac{3}{8} - \frac{\pi^2}{6}\right] \quad (7)$$

where $(\Delta E/E)_{accept}$ is the rf acceptance of the bucket.

Another important limitation, namely, the Touschek beam lifetime in ring is expected to be small, because of the extremely small beam emittance. In order to estimate the Touscheck beam lifetime, following formula can be used [20]:

$$\frac{1}{N}\frac{dN}{dt} = \frac{1}{\tau} = \frac{Nr_0^2 c}{8\pi\sigma_x\sigma_y\sigma_z}\frac{\lambda^3}{\gamma^2}D(\xi) \quad (8)$$

where $\lambda$ is the momentum acceptance, $\sigma_{x,y,z}$ are rms beam sizes in three planes, $\gamma$ is the Lorentz factor, and $\xi = \left(\frac{\Delta E/E}{\gamma}\right)^2 \frac{\beta_x}{\varepsilon_x}$. Bruck's approximation valid for $\xi < 0.01$ is used for the function $D(\xi)$

$$D(\xi) = \sqrt{\xi}\left(\ln\left(\frac{1}{1.78\xi}\right) - \frac{3}{2}\right) \quad (9)$$

## III. NOMINAL PARAMETERS FOR SUPER CHARM FACTORY

In Table I, we present set of parameters for electron and positron beams. Luminosity value is calculated by CAIN simulation program [21] and obtained as $L = 1.4 \cdot 10^{35} \text{cm}^{-2}\text{s}^{-1}$.

Very short bunches are one of the key requirements in high luminosity colliders as this permits a decreased $\beta_y^*$ at the IP, thus increasing the luminosity. In fact, $\beta_y^*$ cannot be made much smaller than the bunch length due to "hourglass" effect. In order to minimize the beam-beam effect, small vertical emittance, together with large horizontal beam size and horizontal emittance is needed for high luminosity. But, in a ring shortening the bunch length $\sigma_z$ is very difficult. One can solve this problem with the proposed crabbed waist scheme [22] for beam-beam collisions. This scheme will be used at superb [9] and BINP tau-charm factory [14]. In our case, taking $\beta_y = 0.3$ mm and $\varepsilon_y^N = 0.06$ μm for ring and $\varepsilon_y^N = 0.02$ for linac in table 1, luminosity value $L = 10^{36} \text{cm}^{-2}\text{s}^{-1}$ can be obtained with crab waist scheme. New tune shifts are $\xi_x / \xi_y = 0.012 / 0.08$ and beam sizes are $\sigma_x = 36$ μm and $\sigma_y = 0.05$ μm.

Table I. ERL-ring electron-positron collider parameters.

| Parameters | Positron ring |
|---|---|
| Positron beam energy $E_{e^+}$ (GeV) | 3.56 |
| Number of positrons per bunch ($10^{11}$) | 2 |
| Beta functions at IP $\beta_x / \beta_y$ (mm) | 80/5 |
| Normalized emittances $\varepsilon_x^N / \varepsilon_y^N$ (μm) | 110/0.36 |
| $\sigma_x / \sigma_y$ (μm) | 36/0.5 |
| $\sigma_z$ (mm) | 5 |
| Beam-beam tune shift $(\xi_x / \xi_y)$ | 0.012/0.13 |
| Energy loss / turn (MeV) | 1.7 |
| Number of buches, $n_b$ | 125 |
| Revolution frequency (MHz) | 1.2 |
| Circumference, C (m) | 250 |
| Beam current (A) | 4.8 |
| Momentum Acceptance (%) | 1 |
| **Parameters** | **Electron ERL** |
| Electron beam energy $E_{e^-}$ (GeV) | 1 |
| Number of electrons per bunch ($10^{10}$) | 2 |
| Beta functions at IP $\beta_x / \beta_y$ (mm) | 80/5 |
| Normalized emittances $\varepsilon_x^N / \varepsilon_y^N$ (μm) | 31/0.1 |
| $\sigma_x / \sigma_y$ (μm) | 36/0.5 |
| $\sigma_z$ (mm) | 5 |
| Beam current (A) | 0.48 |
| **Collider Parameters** | |
| Crossing angle θ (mrad) | 34 |
| Collision frequency (MHz) | 150 |
| Luminosty (cm$^{-2}$s$^{-1}$) | $1.4 \cdot 10^{35}$ |

Table II. Radiative Bhabha cross section and life times for positron ring

|  | Choice I |
|---|---|
| $\sigma_{radiative}$ (mbarn) | 238 |
| Luminosity lifetime (min) | 17.5 |
| Touschek lifetime (min) | 94 |
| Total beam lifetime (min) | 14.8 |

With parameters set given in Table 1, we obtain beam lifetimes presented in Table II (rms beta functions in ring are taken as $\beta_x = 15\,\text{m}$, $\beta_y = 25\,\text{m}$). For crab waist case, total beam lifetime is very small. This problem can be solved by increasing the circumference of the positron ring. For example, with increasing ring circumference from 250 m to 2250 m total beam lifetime is increased from ~1.8 min to ~14 min. which is comparable with that of Super-B factory proposals [9].

## IV. PHYSICS POTENTIAL

As mentioned in section II, in order to utilize high luminosity advantage, it is important to obey condition $\Delta(\sqrt{s}) < \Gamma$. Luminosity spectrum $dL/dW_{cm}$ with the beam parameters given in Table I is plotted in figure 1. CAIN program [21] is used for simulation with $\Delta E_{e^+}/E_{e^+} = \Delta E_{e^-}/E_{e^-} = 10^{-3}$. It is seen that center of mass energy spread is well below $\Gamma_{\psi(3S)} \approx 27\,\text{MeV}$. Therefore, we can use the well known Breit-Wigner formula

$$\sigma_{BW} = \frac{12\pi}{(m_{\Psi(3S)}^2)} B_{in} B_{out} \qquad (11)$$

where $B_{in}$ and $B_{out}$ are the branching fractions of the resonance into the entrance and exit channels. With $Br(\psi(3S) \to e^+e^-) \approx 10^{-5}$ [17] and $L = 1.4 \cdot 10^{35} \text{cm}^{-2}\text{s}^{-1}$, expected number of $\Psi(3S)$ per working year ($10^7$ s) is $1.5 \cdot 10^{10}$.

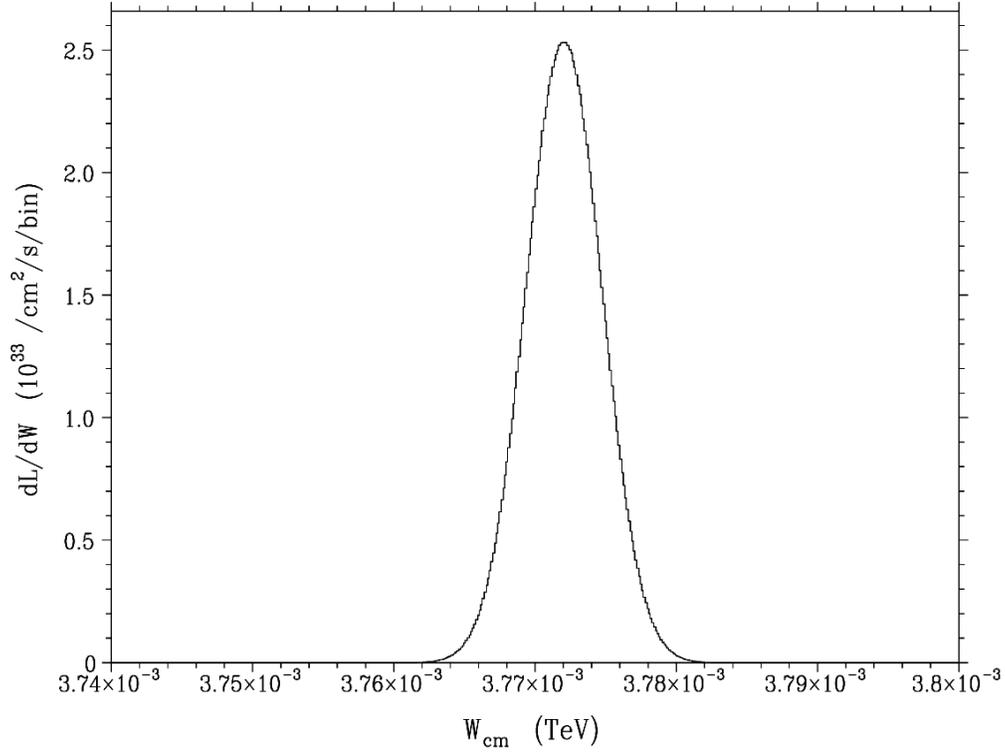

Figure 1. Luminosity Spectrum for Charm Factory

As shown in [16], concerning $D^0 - \overline{D}^0$ mixing and rare charm decay, even with $L = 10^{34} \text{cm}^{-2}\text{s}^{-1}$ dedicated charm factory is advantageous comparing to super-B. Therefore, ERL-ring type charm factory with $L = 1.4 \cdot 10^{35} \text{cm}^{-2}\text{s}^{-1}$ has a great potential to investigated the charm physics. For example, $Br(D^0 \to \mu^+ e^-)$ can be measured up to $10^{-10}$ improving existing experimental upper limit by more than three orders.

## V. CONCLUSIONS

There are two charm factory proposals with luminosity exceeding $L = 10^{35} \text{cm}^{-2}\text{s}^{-1}$. The BINP proposal is a traditional (ring-ring) type one, and high luminosity is achieved due to crab waist collision scheme. Our proposal is less conventional and need essential R&D efforts. Nevertheless, same luminosity is achievable without crab waist collision scheme. If this scheme is used for ERL-ring charm factory a luminosity of $L = 10^{36} \text{cm}^{-2}\text{s}^{-1}$ seems to be possible. This leads to an obvious advantage in search for rare decays. Another important feature of linac-ring type charm factory is the asymmetric kinematics. This will be important in investigation of oscillations and CP-violation in the charm sector of the SM.

## ACKNOWLEDGEMENTS

This work is supported by Turkish Atomic Energy Authority (TAEK) and State Planning Organization (DPT) with grant number DPT2006K-120470.